\documentclass[reprint,aps,prl,linenumbers]{revtex4-1}

\usepackage{amssymb}
\usepackage{amsmath}
\usepackage{array}
\usepackage{dcolumn}
\usepackage{hyperref}
\usepackage{graphicx}
\usepackage{longtable}
\usepackage{footnote}

\begin{document}
\title{Re-publication of the data from the BILL magnetic spectrometer: The cumulative $\beta$ spectra of the fission products of $^{235}$U, $^{239}$Pu, and $^{241}$Pu}
\date{ \today}

\author{N. Haag}
\email[Corresponding author: ]{nils.haag@ph.tum.de}

\author{F.\,von\,Feilitzsch}
\author{L. Oberauer}
\author{W. Potzel}
\author{K. Schreckenbach}
\thanks{deceased}
\affiliation{Physik Department, Technische Universit\"{a}t M\"{u}nchen, 85748 Garching, Germany}
\author{A. A. Sonzogni}
\affiliation{National Nuclear Data Center, Building 817, Brookhaven National Laboratory, Upton, NY 11973-5000, USA}
\author{W. Gelletly}
\affiliation{Physics Department, University of Surrey, Guildford, Surrey, U.K.}

\begin{abstract}
In the 1980s, measurements of the cumulative $\beta$ spectra of the fission products following the thermal neutron induced fission of $^{235}$U, $^{239}$Pu, and $^{241}$Pu were performed at the magnetic spectrometer BILL at the ILL in Grenoble. These data were published in bins of 250\,keV. In this paper, we re-publish the original data with a binning of 50\,keV for $^{235}$U and 100\,keV for $^{239}$Pu and $^{241}$Pu.
\end{abstract}

\maketitle
\subsection{Introduction}
Nuclear reactors are intense emitters of electron-antineutrinos and, thus, are valuable sources used in neutrino physics. In 2012, three major reactor $\bar\nu$ oscillation experiments, Daya Bay~\cite{dayabay}, RENO~\cite{reno}, and Double Chooz~\cite{doublechooz} published their independently measured values of the neutrino oscillation parameter $\theta_{13}$. It was also suggested that a reactor $\bar\nu$ anomaly existed, namely that there was possibly a missing contribution for antineutrinos from reactor cores. This started a debate about the hypothetical existence of so-called sterile neutrinos \cite{mention11}. Small-scale experiments in the vicinity of reactor cores are being built to pin down the reactor anomaly and to support another aim of reactor $\bar\nu$ physics: the non-proliferation of weapons-grade plutonium \cite{NonProl}.\\
In all these experiments, detectors are placed near commercial nuclear power plants. In their cores, neutron induced fission of the four main nuclides ($^{235}$U, $^{238}$U, $^{239}$Pu and $^{241}$Pu) takes place and the $\bar\nu$'s are generated in the $\beta^{-}$ decay of the neutron-rich fission products. The relative contributions of the main fuel isotopes depend on the design of the reactor and the time it has been operating. The total $\bar\nu$ spectrum is a sum of more than $\sim$\,10000 individual beta spectra from over $\sim$\,800 nuclei, weighted with the corresponding transition intensities and fission yields \cite{Mueller}. One must point out that our current knowledge of the fission yields, intensities and shapes data is insufficient to calculate the $\bar\nu$ spectra precisely \cite{fallot12, Hayes}. However, the total $\bar\nu$ spectrum can be derived from measured cumulative $\beta$ spectra of the main fuel isotopes \cite{huber11,Mueller, vogel81}, making accurate measurements of these $\beta$ spectra of the fission products very valuable.\\

The goal of this article is the publication of high-resolution electron spectra recorded at the Institut Laue-Langevin ILL in Grenoble, France, with a finer binning than those originally published in \cite{BILL81,BILL82,BILL85,BILL89}.
\subsection{The data of the BILL experiment}

In the 1980s, target foils of $^{235}$U, $^{239}$Pu, and $^{241}$Pu were irradiated with thermal neutrons at the ILL reactor and the magnetic spectrometer BILL \cite{Mampe} was used to measure the cumulative $\beta$ spectra emitted by the fission products of the irradiated isotopes with high precision \cite{BILL81,BILL82,BILL85,BILL89}. These beta spectra were subsequently converted into the corresponding $\bar\nu_e$ spectra and used in the predictions of the $\bar\nu_e$ spectra emitted by nuclear reactors. Recently, the cumulative $\beta$ spectrum of the fission product of $^{238}$U was also measured at the scientific neutron source FRM\,II in Garching \cite{MyPRL}.\\
The data from the BILL experiment was published with a bin-width of 250\,keV, which was sufficiently accurate for applications at that time. However, the original data were recorded in a finer binning and, thus, spectra with a bin-width of 50\,keV for $^{235}$U and 100\,keV for $^{239}$Pu and $^{241}$Pu are available. Within the present paper, we make these data available to the community.
We do not perform any re-evaluation of the BILL data and refer the reader to the original papers for detailed information about the BILL spectrometer and the data analysis \cite{BILL81,BILL82,BILL85,BILL89}. All errors are given at 1\,$\sigma$ confidence level (68\,\%).

\subsection{The cumulative $\beta$ spectrum of $^{235}$U}
The $\beta$ spectrum of the fission products of $^{235}$U is available in 50\,keV\,-\,wide bins and is given in table \ref{Tab:U235new}. In addition to the bin-to-bin uncorrelated statistical error, an (only slightly energy dependent) bin-to-bin correlated error on the absolute rate is 1.7\% at 1.3\,MeV and 1.9\,\% at 7.4\,MeV \cite{BILL85}. The error may be linearly interpolated for other energies.\\
To minimize off-equilibrium effects, only data taken $>$\,12\,h after the start of the irradiation of the target foil were used at energies up to 5\,MeV.

\subsection{The cumulative $\beta$ spectrum of $^{239}$Pu}
The spectrum of the fission products of $^{239}$Pu is available in 100\,keV\,-\,wide bins and is given in table \ref{Tab:Pu239new}. The uncertainties of the data points were calculated (as instructed by the authors of the original paper) from the original 250\,keV data \cite{BILL82}:
\begin{equation}\label{Eq:Pu239Err}
 \sigma(E)_{100} = \sqrt{2.5} \cdot \sigma(E)_{250}
\end{equation}
Herein, $\sigma(E)_{100}$ is the error calculated for the 100\,keV data point at energy E and $\sigma(E)_{250}$ is the uncertainty given in the original data. The factor $\sqrt{2.5}$ statistically takes into account the different number of entries in the differently sized bins.\\
In addition to the statistical error, an error on the absolute normalization of 2.0\% at 2.5\,MeV and 2.6\,\% at 6.5\,MeV has to be taken into account \cite{BILL82}. The error may be linearly interpolated for other energies.\\
Similar to the $^{235}$U measurement, only data taken $>$\,1.5\,d after the start of the irradiation of the target were used in the analysis.

\subsection{The cumulative $\beta$ spectrum of $^{241}$Pu}
Data for $^{241}$Pu are also available in 100\,keV binning (table \ref{Tab:Pu241new}). After publication of \cite{BILL89}, further analysis of the spectra were performed and the errors given, estimated rather conservatively in the original paper, could be slightly reduced. Thus, the errors quoted are smaller than those one would obtain by the procedure applied to the $^{239}$Pu data (see also eq. \ref{Eq:Pu239Err}). The absolute normalization error to be added to this statistical uncertainty is 1.8\,\% at 1.5\,MeV and 1.9\,\% at 7\,MeV \cite{BILL89}.\\
Analogous to the other spectra, to minimize off-equilibrium effects, only data from irradiation times $>$\,1.8\,d were used at energies up to 3.5\,MeV.

\newpage
\section{Appendix}

\begin{longtable}{|c|c|c|}
\caption{\textit{The cumulative $\beta$ spectrum of the fission products of $^{235}$U in units of electrons per fission and MeV. E$_{kin}$ is the kinetic energy of the electrons and indicates the center of the 50\,keV wide energy bins. The statistical error (last column) is given at 68\,\% confidence level.}} \label{Tab:U235new}\\
\hline 
E$_{kin}$ [keV] & N$_{\beta}$ $\left[ \frac{betas}{fission \cdot MeV}\right]$ & stat. err. [\%] \\
\hline
\endfirsthead
\hline
E$_{kin}$ [keV] & N$_{\beta}$ $\left[ \frac{betas}{fission \cdot MeV}\right]$ & stat. err. [\%] \\
\hline
\endhead
\hline
\multicolumn{3}{|c|}{Continued ...}\\
\hline
\endfoot
\hline
\multicolumn{3}{|c|}{End of table}\\
\hline
\endlastfoot

\hline	
1500 & 1.352 	& 0.14\\
1550 & 1.322	 & 0.14\\
1600 & 1.266 &0.14\\
1650 & 1.221 	&0.14\\
1700 & 1.168 	&0.14\\
1750 & 1.134 	&0.14\\
1800 & 1.090 	&0.14\\
1850 & 1.053 	&0.14\\
1900 & 1.006 	&0.14\\
1950 & 9.727 $\times$ 10$^{-1}$&	0.14\\
2000 & 9.319 $\times$ 10$^{-1}$&	0.14\\
2050 & 8.984 $\times$ 10$^{-1}$&	0.14\\
2100 & 8.632 $\times$ 10$^{-1}$&	0.15\\
2150 & 8.241 $\times$ 10$^{-1}$&	0.15\\
2200 & 7.993 $\times$ 10$^{-1}$&	0.15\\
2250 & 7.713 $\times$ 10$^{-1}$&	0.15\\
2300 & 7.443 $\times$ 10$^{-1}$&	0.15\\
2350 & 7.106 $\times$ 10$^{-1}$&	0.15\\
2400 & 6.796 $\times$ 10$^{-1}$	&0.15\\
2450 & 6.584 $\times$ 10$^{-1}$&	0.15\\
2500 & 6.337 $\times$ 10$^{-1}$&	0.16\\
2550 & 6.108 $\times$ 10$^{-1}$&	0.16\\
2600 & 5.868 $\times$ 10$^{-1}$&	0.16\\
2650 & 5.657 $\times$ 10$^{-1}$&	0.16\\
2700 & 5.440 $\times$ 10$^{-1}$&	0.16\\
2750 & 5.221 $\times$ 10$^{-1}$&	0.16\\
2800 & 5.026 $\times$ 10$^{-1}$&	0.17\\
2850 & 4.816 $\times$ 10$^{-1}$&	0.17\\
2900 & 4.577 $\times$ 10$^{-1}$&	0.17\\
2950 & 4.419 $\times$ 10$^{-1}$&	0.17\\
3000 & 4.241 $\times$ 10$^{-1}$&	0.17\\
3050 & 4.057 $\times$ 10$^{-1}$&	0.18\\
3100 & 3.893 $\times$ 10$^{-1}$&	0.18\\
3150 & 3.726 $\times$ 10$^{-1}$&	0.18\\
3200 & 3.547 $\times$ 10$^{-1}$&	0.19\\
3250 & 3.383 $\times$ 10$^{-1}$&	0.19\\
3300 & 3.244 $\times$ 10$^{-1}$&	0.19\\
3350 & 3.088 $\times$ 10$^{-1}$&	0.19\\
3400 & 2.942 $\times$ 10$^{-1}$&	0.20\\
3450 & 2.808 $\times$ 10$^{-1}$&	0.20\\
3500 & 2.678 $\times$ 10$^{-1}$&	0.21\\
3550 & 2.560 $\times$ 10$^{-1}$&	0.21\\
3600 & 2.431 $\times$ 10$^{-1}$&	0.21\\
3650 & 2.314 $\times$ 10$^{-1}$&	0.22\\
3700 & 2.205 $\times$ 10$^{-1}$&	0.22\\
3750 & 2.114 $\times$ 10$^{-1}$&	0.23\\
3800 & 2.015 $\times$ 10$^{-1}$&	0.23\\
3850 & 1.906 $\times$ 10$^{-1}$&	0.24\\
3900 & 1.809 $\times$ 10$^{-1}$&	0.24\\
3950 & 1.721 $\times$ 10$^{-1}$&	0.25\\
4000 & 1.632 $\times$ 10$^{-1}$&	0.25\\
4050 & 1.547 $\times$ 10$^{-1}$&	0.26\\
4100 & 1.467 $\times$ 10$^{-1}$&	0.26\\
4150 & 1.388 $\times$ 10$^{-1}$&	0.27\\
4200 & 1.321 $\times$ 10$^{-1}$&	0.27\\
4250 & 1.252 $\times$ 10$^{-1}$&	0.28\\
4300 & 1.191 $\times$ 10$^{-1}$&	0.29\\
4350 & 1.135 $\times$ 10$^{-1}$&	0.29\\
4400 & 1.077 $\times$ 10$^{-1}$&	0.30\\
4450 & 1.020 $\times$ 10$^{-1}$&	0.31\\
4500 & 9.687 $\times$ 10$^{-2}$&	0.31\\
4550 & 9.153 $\times$ 10$^{-2}$&0.32\\
4600 & 8.747 $\times$ 10$^{-2}$& 0.33\\
4650 & 8.328 $\times$ 10$^{-2}$&	0.33\\
4700 & 7.975 $\times$ 10$^{-2}$	&0.34\\
4750 & 7.576 $\times$ 10$^{-2}$&	0.35\\
4800 & 7.146 $\times$ 10$^{-2}$	&0.36\\
4850 & 6.756 $\times$ 10$^{-2}$&	0.37\\
4900 & 6.386 $\times$ 10$^{-2}$&	0.38\\
4950 & 6.111 $\times$ 10$^{-2}$&	0.38\\
5000 & 5.736 $\times$ 10$^{-2}$&	0.28\\
5050 & 5.474 $\times$ 10$^{-2}$&	0.28\\
5100 & 5.160 $\times$ 10$^{-2}$&	0.29\\
5150 & 4.876 $\times$ 10$^{-2}$&	0.30\\
5200 & 4.614 $\times$ 10$^{-2}$&	0.31\\
5250 & 4.344 $\times$ 10$^{-2}$&	0.32\\
5300 & 4.121 $\times$ 10$^{-2}$&	0.33\\
5350 & 3.912 $\times$ 10$^{-2}$&	0.34\\
5400 & 3.711 $\times$ 10$^{-2}$&	0.35\\
5450 & 3.527 $\times$ 10$^{-2}$&	0.36\\
5500 & 3.315 $\times$ 10$^{-2}$&	0.37\\
5550 & 3.136 $\times$ 10$^{-2}$&	0.38\\
5600 & 2.958 $\times$ 10$^{-2}$&	0.39\\
5650 & 2.773 $\times$ 10$^{-2}$&	0.40\\
5700 & 2.621 $\times$ 10$^{-2}$&	0.42\\
5750 & 2.465 $\times$ 10$^{-2}$&	0.43\\
5800 & 2.323 $\times$ 10$^{-2}$&	0.44\\
5850 & 2.206 $\times$ 10$^{-2}$&	0.45\\
5900 & 2.068 $\times$ 10$^{-2}$&	0.47\\
5950 & 1.923 $\times$ 10$^{-2}$&	0.48\\
6000 & 1.815 $\times$ 10$^{-2}$&	0.49\\
6050 & 1.681 $\times$ 10$^{-2}$&	0.51\\
6100 & 1.591 $\times$ 10$^{-2}$&	0.52\\
6150 & 1.502 $\times$ 10$^{-2}$&	0.53\\
6200 & 1.409 $\times$ 10$^{-2}$&	0.55\\
6250 & 1.316 $\times$ 10$^{-2}$&	0.56\\
6300 & 1.230 $\times$ 10$^{-2}$&	0.58\\
6350 & 1.150 $\times$ 10$^{-2}$&	0.60\\
6400 & 1.068 $\times$ 10$^{-2}$&	0.62\\
6450 & 9.932 $\times$ 10$^{-3}$&	0.64\\
6500 & 9.265 $\times$ 10$^{-3}$&	0.46\\
6550 & 8.587 $\times$ 10$^{-3}$&	0.48\\
6600 & 7.954 $\times$ 10$^{-3}$&	0.50\\
6650 & 7.290 $\times$ 10$^{-3}$&	0.53\\
6700 & 6.678 $\times$ 10$^{-3}$&	0.56\\
6750 & 6.170 $\times$ 10$^{-3}$&	0.58\\
6800 & 5.629 $\times$ 10$^{-3}$&	0.61\\
6850 & 5.135 $\times$ 10$^{-3}$&	0.64\\
6900 & 4.700 $\times$ 10$^{-3}$&	0.68\\
6950 & 4.275 $\times$ 10$^{-3}$&	0.72\\
7000 & 3.832 $\times$ 10$^{-3}$&	0.77\\
7050 & 3.492 $\times$ 10$^{-3}$&	0.81\\
7100 & 3.175 $\times$ 10$^{-3}$&	0.86\\
7150 & 2.86 $\times$ 10$^{-3}$&	0.91\\
7200 & 2.616 $\times$ 10$^{-3}$&	0.95\\
7250 & 2.352 $\times$ 10$^{-3}$&	1.0\\
7300 & 2.088 $\times$ 10$^{-3}$&	1.1\\
7350 & 1.911 $\times$ 10$^{-3}$&	1.1\\
7400 & 1.682 $\times$ 10$^{-3}$&	1.2\\
7450 & 1.473 $\times$ 10$^{-3}$&	1.2\\
7500 & 1.303 $\times$ 10$^{-3}$&	1.3\\
7550 & 1.170 $\times$ 10$^{-3}$&	1.4\\
7600 & 1.018 $\times$ 10$^{-3}$&	1.5\\
7650 & 8.553 $\times$ 10$^{-4}$&	1.7\\
7700 & 7.306 $\times$ 10$^{-4}$&	1.9\\
7750 & 6.111 $\times$ 10$^{-4}$&	2.1\\
7800 & 5.342 $\times$ 10$^{-4}$&	2.4\\
7850 & 4.301 $\times$ 10$^{-4}$&	2.8\\
7900 & 3.442 $\times$ 10$^{-4}$&	3.3\\
7950 & 2.931 $\times$ 10$^{-4}$&	3.7\\
8000 & 2.403 $\times$ 10$^{-4}$&	3.9\\
8050 & 2.032 $\times$ 10$^{-4}$&	4.5\\
8100 & 1.679 $\times$ 10$^{-4}$&	5.3\\
8150 & 1.454 $\times$ 10$^{-4}$&	5.9\\
8200 & 1.3 $\times$ 10$^{-4}$	&6.5\\
8250 & 1.259 $\times$ 10$^{-4}$	&6.4\\
8300 & 1.14 $\times$ 10$^{-4}$	&6.7\\
8350 & 1.015 $\times$ 10$^{-4}$	&7.0\\
8400 & 8.41 $\times$ 10$^{-5}$	&8.2\\
8450 & 7.32 $\times$ 10$^{-5}$	&9.3\\
8500 & 5.63 $\times$ 10$^{-5}$	&12\\
8550 & 5.51 $\times$ 10$^{-5}$	&12\\
8600 & 4.14 $\times$ 10$^{-5}$	&15\\
8650 & 3.96 $\times$ 10$^{-5}$	&15\\
8700 & 3.56 $\times$ 10$^{-5}$	&17\\
8750 & 3.11 $\times$ 10$^{-5}$	&16\\
8800 & 2.14 $\times$ 10$^{-5}$	&21\\
8850 & 1.52 $\times$ 10$^{-5}$	&29\\
8900 & 1.43 $\times$ 10$^{-5}$	&31\\
8950 & 1.16 $\times$ 10$^{-5}$	&38\\
9000 & 1.42 $\times$ 10$^{-5}$	&31\\
9050 & 1.03 $\times$ 10$^{-5}$	&38\\
9100 & 8.4 $\times$ 10$^{-6}$	&45\\
9150 & 1.42 $\times$ 10$^{-5}$	&27\\
9200 & 1.39 $\times$ 10$^{-5}$	&27\\
9250 & 9.2 $\times$ 10$^{-6}$	&40\\
9300 & 6 $\times$ 10$^{-6}$	&58\\
9350 & 8.6 $\times$ 10$^{-6}$	&41\\
9400 & 5.9 $\times$ 10$^{-6}$	&58\\
9450 & 1 $\times$ 10$^{-6}$	&3.3 $\times$ 10$^{2}$\\
9500 & 9.7 $\times$ 10$^{-6}$	&33\\
9550 & 4.7 $\times$ 10$^{-6}$	&68\\
9600 & 2.4 $\times$ 10$^{-6}$	&1.3 $\times$ 10$^{2}$\\
	\hline
	\hline	
\end{longtable}

\begin{longtable}{|c|c|c|}

\caption{\textit{The cumulative $\beta$ spectrum of the fission products of $^{239}$Pu in units of electrons per fission and MeV. E$_{kin}$ is the kinetic energy of the electrons and indicates the center of the 100\,keV wide energy bins. The statistical error (last column) is given at 68\,\% confidence level.}} \label{Tab:Pu239new}\\
\hline 
E$_{kin}$ [keV] & N$_{\beta}$ $\left[ \frac{betas}{fission \times MeV}\right]$ & stat. err. [\%] \\
\hline
\endfirsthead
\hline
E$_{kin}$ [keV] & N$_{\beta}$ $\left[ \frac{betas}{fission \times MeV}\right]$ & stat. err. [\%] \\
\hline
\endhead
\hline
\multicolumn{3}{|c|}{Continued ...}\\
\hline
\endfoot
\hline
\multicolumn{3}{|c|}{End of table}\\
\hline
\endlastfoot

\hline	
1500 &	1.14		&	$<$0.79\\
1600 & 	1.06		&\\
1700	& 9.65 $\times$ 10$^{-1}$	&\\
1800	&8.86 $\times$ 10$^{-1}$	&\\
1900	&8.16 $\times$ 10$^{-1}$	&\\
2000	&7.46 $\times$ 10$^{-1}$	&$<$0.79\\
2100	&6.88 $\times$ 10$^{-1}$	&\\
2200	&6.27 $\times$ 10$^{-1}$	&\\
2300	&5.73 $\times$ 10$^{-1}$	&\\
2400	&5.25 $\times$ 10$^{-1}$	&\\
2500	&4.85 $\times$ 10$^{-1}$	&$<$0.79\\
2600	&4.44	$\times$ 10$^{-1}$&\\
2700	&4.05	$\times$ 10$^{-1}$&\\
2800	&3.69	$\times$ 10$^{-1}$&\\
2900	&3.37	$\times$ 10$^{-1}$&\\
3000	&3.04	$\times$ 10$^{-1}$&$<$0.79\\
3100	&2.77	$\times$ 10$^{-1}$&\\
3200	&2.48	$\times$ 10$^{-1}$&\\
3300	&2.24	$\times$ 10$^{-1}$&\\
3400	&2.02	$\times$ 10$^{-1}$&\\
3500	&1.83	$\times$ 10$^{-1}$&0.79\\
3600	&1.62	$\times$ 10$^{-1}$&\\
3700	&1.43	$\times$ 10$^{-1}$&\\
3800	&1.29	$\times$ 10$^{-1}$&\\
3900	&1.16	$\times$ 10$^{-1}$&\\
4000	&1.03	$\times$ 10$^{-1}$&0.95\\
4100	&8.99	$\times$ 10$^{-2}$&\\
4200	&7.99	$\times$ 10$^{-2}$&\\
4300	&6.95	$\times$ 10$^{-2}$&\\
4400	&6.20	$\times$ 10$^{-2}$&\\
4500	&5.50	$\times$ 10$^{-2}$&1.3\\
4600	&4.86	$\times$ 10$^{-2}$&\\
4700	&4.36	$\times$ 10$^{-2}$&\\
4800	&3.85	$\times$ 10$^{-2}$&\\
4900	&3.49	$\times$ 10$^{-2}$&\\
5000	&3.17	$\times$ 10$^{-2}$&1.6\\
5100	&2.76	$\times$ 10$^{-2}$&\\
5200	&2.43	$\times$ 10$^{-2}$&\\
5300	&2.19	$\times$ 10$^{-2}$&\\
5400	&1.93	$\times$ 10$^{-2}$&\\
5500	&1.73	$\times$ 10$^{-2}$&2.4\\
5600	&1.49	$\times$ 10$^{-2}$&\\
5700	&1.27	$\times$ 10$^{-2}$&\\
5800	&1.08	$\times$ 10$^{-2}$&\\
5900	&9.52	$\times$ 10$^{-3}$&\\
6000	&8.28	$\times$ 10$^{-3}$&3.6\\
6100	&7.36	$\times$ 10$^{-3}$&\\
6200	&6.44	$\times$ 10$^{-3}$&\\
6300	&5.49	$\times$ 10$^{-3}$&\\
6400	&4.9$\times$ 10$^{-3}$	&\\
6500	&4.21	$\times$ 10$^{-3}$&4.3\\
6600	&3.54	$\times$ 10$^{-3}$&\\
6700	&2.77	$\times$ 10$^{-3}$&\\
6800	&2.37	$\times$ 10$^{-3}$&\\
6900	&1.98	$\times$ 10$^{-3}$&\\
7000	&1.68	$\times$ 10$^{-3}$&8.7\\
7100	&1.44	$\times$ 10$^{-3}$&\\
7200	&1	$\times$ 10$^{-3}$&\\
7300	&8.1	$\times$ 10$^{-4}$&\\
7400	&6.5	$\times$ 10$^{-4}$&\\
7500	&5$\times$ 10$^{-4}$	&16\\
7600	&4	$\times$ 10$^{-4}$&\\
7700	&3.4$\times$ 10$^{-4}$	&\\
7800	&2.9$\times$ 10$^{-4}$	&\\
7900	&2.5$\times$ 10$^{-4}$	&\\
8000	&2	$\times$ 10$^{-4}$&32\\
\hline
\hline
\end{longtable}

\begin{longtable}{|c|c|c|}

\caption{\textit{The cumulative $\beta$ spectrum of the fission products of $^{241}$Pu in units of electrons per fission and MeV. E$_{kin}$ is the kinetic energy of the electrons and indicates the center of the 100\,keV wide energy bins. The statistical error (last column) is given at 68\,\% confidence level.}} \label{Tab:Pu241new}\\
\hline 
E$_{kin}$ [keV] & N$_{\beta}$ $\left[ \frac{betas}{fission \times MeV}\right]$ & stat. err. [\%] \\
\hline
\endfirsthead
\hline
E$_{kin}$ [keV] & N$_{\beta}$ $\left[ \frac{betas}{fission \times MeV}\right]$ & stat. err. [\%] \\
\hline
\endhead
\hline
\multicolumn{3}{|c|}{Continued ...}\\
\hline
\endfoot
\hline
\multicolumn{3}{|c|}{End of table}\\
\hline
\endlastfoot

\hline	
1500	&1.31	&$<$\,0.8\\
1600	&1.22	&\\
1700	&1.14	&\\
1800	&1.06	&\\
1900	&9.84	$\times$ 10$^{-1}$&\\
2000	&9.11	$\times$ 10$^{-1}$&\\
2100	&8.43	$\times$ 10$^{-1}$	&\\
2200	&7.80		$\times$ 10$^{-1}$&\\
2300	&7.23		$\times$ 10$^{-1}$&\\
2400	&6.67		$\times$ 10$^{-1}$&\\
2500	&6.19	$\times$ 10$^{-1}$&\\
2600	&5.72		$\times$ 10$^{-1}$&\\
2700	&5.26		$\times$ 10$^{-1}$&\\
2800	&4.83		$\times$ 10$^{-1}$&\\
2900	&4.43		$\times$ 10$^{-1}$&\\
3000	&4.06	$\times$ 10$^{-1}$	&0.8\\
3100	&3.71		$\times$ 10$^{-1}$&\\
3200	&3.37		$\times$ 10$^{-1}$&\\
3300	&3.07		$\times$ 10$^{-1}$&\\
3400	&2.79		$\times$ 10$^{-1}$&\\
3500	&2.52		$\times$ 10$^{-1}$&\\
3600	&2.28	$\times$ 10$^{-1}$	&\\
3700	&2.06	$\times$ 10$^{-1}$	&\\
3800	&1.86		$\times$ 10$^{-1}$&\\
3900	&1.67	$\times$ 10$^{-1}$	&\\
4000	&1.49		$\times$ 10$^{-1}$&1\\
4100	&1.33	$\times$ 10$^{-1}$	&\\
4200	&1.18		$\times$ 10$^{-1}$&\\
4300	&1.06		$\times$ 10$^{-1}$&\\
4400	&9.40	$\times$ 10$^{-2}$	&\\
4500	&8.41	$\times$ 10$^{-2}$&\\
4600	&7.55		$\times$ 10$^{-2}$&\\
4700	&6.75		$\times$ 10$^{-2}$&\\
4800	&6.02		$\times$ 10$^{-2}$&\\
4900	&5.36 	$\times$ 10$^{-2}$	&\\
5000	&4.73 	$\times$ 10$^{-2}$	&1.6\\
5100	&4.20		$\times$ 10$^{-2}$&\\
5200	&3.70		$\times$ 10$^{-2}$&\\
5300	&3.26	 $\times$ 10$^{-2}$	&\\
5400	&2.86		$\times$ 10$^{-2}$&\\
5500	&2.50		$\times$ 10$^{-2}$&\\
5600	&2.18		$\times$ 10$^{-2}$&\\
5700	&1.89		$\times$ 10$^{-2}$&\\
5800	&1.65		$\times$ 10$^{-2}$&\\
5900	&1.43		$\times$ 10$^{-2}$&\\
6000	&1.24		$\times$ 10$^{-2}$&2.5\\
6100	&1.08		$\times$ 10$^{-2}$&\\
6200	&9.39		$\times$ 10$^{-3}$&\\
6300	&8.18			$\times$ 10$^{-3}$&\\
6400	&7.06			$\times$ 10$^{-3}$&\\
6500	&5.99			$\times$ 10$^{-3}$&\\
6600	&5.07			$\times$ 10$^{-3}$&\\
6700	&4.24			$\times$ 10$^{-3}$&\\
6800	&3.54			$\times$ 10$^{-3}$&\\
6900	&2.90			$\times$ 10$^{-3}$&\\
7000	&2.34			$\times$ 10$^{-3}$&4.4\\
7100	&1.92			$\times$ 10$^{-3}$&\\
7200	&1.59			$\times$ 10$^{-3}$&\\
7300	&1.28			$\times$ 10$^{-3}$&\\
7400	&1.05			$\times$ 10$^{-3}$&\\
7500	&8.26			$\times$ 10$^{-4}$&	5\\
7600	&6.70		$\times$ 10$^{-4}$&\\
7700	&5.15		$\times$ 10$^{-4}$&\\
7800	&4.04		$\times$ 10$^{-4}$&\\
7900	&2.95		$\times$ 10$^{-4}$&\\
8000	&2.32		$\times$ 10$^{-4}$&	9\\
8100	&1.68		$\times$ 10$^{-4}$&\\
8200	&1.34		$\times$ 10$^{-4}$&\\
8300	&1.03		$\times$ 10$^{-4}$&\\
8400	&7.02		$\times$ 10$^{-5}$&\\
8500	&5.59		$\times$ 10$^{-5}$ &18\\
8600	&4.23		$\times$ 10$^{-5}$&\\
8700	&2.90		$\times$ 10$^{-5}$&\\
8800	&1.68		$\times$ 10$^{-5}$&\\
8900	&1.16		$\times$ 10$^{-5}$&\\
9000	&6.29 		$\times$ 10$^{-6}$&73\\
\hline
\hline
\end{longtable}


\begin{thebibliography}{xxxxxxx}
 	\addcontentsline{toc}{chapter}{Bibliography}

	\bibitem{dayabay} F.\,P. An et al., Phys. Rev. Lett. 108, 171803 (2012)                     
	\bibitem{reno} J.\,K. Ahn et al., Phys. Rev. Lett. 108, 191802 (2012)                   
	\bibitem{doublechooz} Y. Abe et al., Phys. Rev. Lett. 108, 131801 (2012)
	\bibitem{mention11} G. Mention et al., Phys. Rev. D 83, 073006 (2011)
	\bibitem{NonProl}H. Furuta, H. Tadokoro, A. Imura, Y. Furuta, T. Niisato, and F. Suekane, Report No. IAEA-CN-184/63, (2010) (unpublished)
	\bibitem{Mueller} Th.\,A. Mueller et al., Phys. Rev. C 83, 054615 (2011)
	\bibitem{fallot12} M. Fallot et al., Phys. Rev. Lett. 109, 202504 (2012)
	\bibitem{Hayes} A.\,C. Hayes, J.\,L. Friar, G.\,T. Garvey, G. Jonkmans, arXiv:1309.4146 (2013)
	\bibitem{vogel81} P. Vogel et al., Phys. Rev. C 24, 1543 (1981)
	\bibitem{huber11} P. Huber, Phys. Rev. C 84, 024617 (2011)           
	\bibitem{BILL81} K. Schreckenbach, H.\,R. Faust, F. von Feilitzsch, A.\,A. Hahn, K. Hawerkamp, and J.\,L. Vuilleumier, Phys. Lett. 99\,B, 251 (1981)
	\bibitem{BILL82} F. von Feilitzsch, A.\,A. Hahn, and K. Schreckenbach, Phys. Lett. 118\,B, 162 (1982)
	\bibitem{BILL85} K. Schreckenbach, G. Covlin, W. Gelletly, and F. von Feilitzsch, Phys. Lett. 160\,B, 325 (1985)
	\bibitem{BILL89} A.\,A. Hahn, K. Schreckenbach, W. Gelletly, F. von Feilitzsch, G. Colvin, and B. Krusche, Phys. Lett. B\,218, 365 (1989)
	\bibitem{Mampe} W. Mampe, K. Schreckenbach, P. Jeuch, B.\,P.\,K. Maier, F. Braumandl, J. Larysz, T. von Egidy, Nucl. Instrum. Meth. 154, 127 (1978)
	\bibitem{MyPRL} N. Haag, A. G\"utlein, M. Hofmann, L. Oberauer, W. Potzel, K. Schreckenbach, and F.\,M. Wagner, Phys. Rev. Lett. 112, 122501 (2014) and N. Haag, PhD Thesis, Technische Universit\"at M\"unchen (2013).
\end{thebibliography}
\end{document}